\documentstyle[12pt,amssymb]{article}

\def\b{\begin{eqnarray}}
\def\e{\end{eqnarray}}
\def\l{\label}
\def\ep{$e^+ e^-$ }
\def\l{\label}
\def\a{\alpha}
\def\be{\beta}

\def\n{(\nu,\tilde\nu)}

\title{
\begin{flushright}
{\large Yaroslavl State University \\
        Preprint YARU-HE-99/06 \\
        astro-ph/9909154} \\[10mm]
\end{flushright}
On the Possible Enhancement of the Magnetic Field \\ 
by Neutrino Reemission Processes \\ 
in the Mantle of a Supernova}

\author{A.A.~Gvozdev\footnote{E-mail: \quad gvozdev@uniyar.ac.ru}
and I.S.~Ognev \\[2mm]
{\it Yaroslavl State (Demidov) University, 
     150000 Yaroslavl, Russia} \\[2mm]
Pis'ma Zh. Eksp. Teor. Fiz. {\bf 69}, 337-342 (1999) \\  
$[$JETP Letters {\bf 69}, 365-370 (1999)$]$}

\date{}

\begin{document}

\large

\maketitle

%\vspace{-16mm}

\begin{abstract}

\baselineskip=18pt

{\normalsize
URCA neutrino reemission processes under the conditions in the
mantle of a supernova with a strong toroidal magnetic field are
investigated. It is shown that parity violation in these processes
can be manifested macroscopically as a torque that rapidly spins up
the region of the mantle occupied by such a field. Neutrino spin-up
of the mantle can strongly affect the mechanism of further
generation of the toroidal field, specifically, it can enhance the
field in a small neighborhood of the rigid-body-rotating core of the
supernova remnant.}

\end{abstract}

%\hspace{3mm}


\vspace*{10mm}

The problem of the shedding of the mantle in an explosion of a
type-II supernova is still far from a complete solution~\cite{R}. It
is known that in several seconds after the collapse of a presupernova
an anomalously high neutrino flux with typical luminosities $L\sim 10^{52}$
ergs/s is emitted from the neutrinosphere, which is approximately
of the same size as the remnant core~\cite{Imshenik}.
In principle, such a neutrino flux could initiate a process leading
to the shedding of the mantle as a result of the absorption and scattering 
of neutrinos by nucleons and the \ep plasma of the medium~\cite{Colgate}. 
However, detailed calculations in spherically
symmetric collapse models have shown that such processes are too weak
for mantle shedding~\cite{N}. In a magnetorotational model~\cite{BK1},
mantle shedding is initiated by the outward pressure of a strong
toroidal magnetic field generated by the differential rotation of the
mantle with the core's primary poloidal magnetic field "frozen" into it.
Indeed, as calculations show~\cite{Bisnovat}, when a mantle rotates with 
a millisecond period in a poloidal field $B_0 \sim 10^{12} - 10^{13}$~G  
a toroidal field $B \sim 10^{15} - 10^{16}$~G is generated in a time of 
order of a second. We note that the model in Ref.~\cite{BK1} 
contains a fundamental limitation on the energy of the toroidal field (it
cannot exceed the kinetic energy of the "core + shell" system) and
therefore on the maximum field itself.

In the present letter we investigate the possibility that this magnetic
field "frozen in" the mantle is enhanced as the result of elementary neutrino
reemission processes occurring in the mantle. We assume the mantle in the
vicinity of the neutrinosphere to be a hot ($T \sim$ several MeV) and
quite dense (though transparent to neutrinos, 
$\rho \sim 10^{11} - 10^{12}$~g/cm$^3$) medium consisting of free
nucleons and \ep plasma. Under these conditions the dominant neutrino
reemission processes are the URCA processes:
\b
p+e^- \rightarrow n + \nu_e \; ,\l{1} \\
n+e^+ \rightarrow p + \tilde \nu_e \; ,\l{2} \\
n +\nu_e\rightarrow p+e^- \; ,\l{3}\\
p + \tilde \nu_e\rightarrow n+e^+ \; .\l{4}
\e
We note that $\be$ decay is statistically suppressed in such a medium. 
The basic idea of this letter is as follows. In an external magnetic
field, neutrinos are emitted and absorbed  asymmetrically with respect to
the direction of the magnetic field as a result of the parity violation 
in the processes (\ref{1})--(\ref{4})~\cite{Chugai}. Therefore a
macroscopic torque spinning up the mantle can arise in a toroidal field.
It is known~\cite{Lai} that for an equilibrium neutrino distribution
function such a neutrino-recoil momentum must be zero. However, the
supernova region under consideration is nonequilibrium for neutrinos, so
that the torque that arises in it is different from zero. Moreover, as
we shall show below, the torque can be large enough to change substantially
the distribution of the angular rotational velocities of the mantle in
the region filled with a strong magnetic field over the characteristic
neutrino emission times. According to the equation governing the
generation of a toroidal field \cite{Bisnovat}, a large change in the
gradient of the angular velocities in the region can lead to
redistribution of the magnetic field (specifically, enhancement of the
field in a small neighborhood of the rigid-body-rotating core).

A quantitative estimate of the effect follows from the expression for the
energy-momentum transferred by neutrinos to a unit volume of the mantle
per unit time:  
\b
\frac{dP_\a}{dt} = \frac{1}{V} \int \prod\limits_i dn_i f_i
\prod\limits_f dn_f (1 - f_f) \frac{|S_{if}|^2}{{\cal T}} k_\a , \l{dpa}
\e
where $dn_i$ and $dn_f$ are the number of initial and final states in an 
element of the phase space, $f_i$ and $f_f$ are the distribution functions 
of the initial and final particles, $k_\a$ is the neutrino momentum, 
$|S_{if}|^2 / {\cal T}$ is the squared $S$-matrix element per unit time. 
It is of interest to calculate the latter under our conditions, since, 
as far as we know, the URCA processes (\ref{1})--(\ref{4}) have been 
previously studied for relatively weak~\cite{Dorofeev} ($B \lesssim m_e^2 / e$) 
and very strong~\cite{Leinson} ($B \sim m_p^2 / e$) fields. Assuming that 
electrons and positrons in the plasma mainly occupy only the lowest Landau 
level ($ \mu_e \lesssim \sqrt{2eB}$, where $\mu_e$ is the chemical potential 
of electrons), we obtained the following expression for the squared 
$S$-matrix element summed over all proton Landau levels and polarizations 
of the final particles and averaged over the polarizations of the initial 
particles:
\b
|S_{if}|^2 & = & 
\frac{G_F^2 \cos^2 \theta_c (2\pi)^3 \; {\cal T}}{2 L_y L_z V^2} \, 
\frac{\exp(-Q_\perp^2 / 2eB)}{4 \omega \varepsilon} \l{s2} \\
& \times & \frac{1}{2} \, 
\left [
\sum\limits_{n=0}^{\infty} \frac{|M_+|^2}{n!}
\left( \frac{Q_\perp^2}{2eB}
\right)^n  \delta^{(3)}
+ \sum\limits_{n=1}^{\infty} \frac{|M_-|^2}{(n-1)!}
\left( \frac{Q_\perp^2}{2eB}
\right)^{n-1}  \delta^{(3)}
\right], 
\nonumber
\e
\b
|M_\sigma|^2 & = & 4 (1 + g_a \sigma)^2
\Bigg[ 2(up)(uk) - (pk) - (up)(k \tilde\varphi u) -
(uk)(p \tilde\varphi u) \Bigg] 
\l{m2} \\
& + & 8 g_a^2 (1 + \sigma)
\Bigg[ (pk) - (p \tilde\varphi k) \Bigg] \nonumber ,
\e
where $u_\a$ is a four-velosity of a medium,   
${\bf B} = (0,0,B)$, and $\delta^{(3)}$ is the production of the    
energy delta-function and two delta-functions of the momentum in the 
direction of the magnetic field and one transverse momentum component 
which are conserved in the reactions; $(pk) = \varepsilon \omega - p_3 k_3$, 
$p^\a = (\varepsilon, {\bf p})$ and $k^\a = (\omega, {\bf k})$ are 
the four-momenta of the electron and neutrino, $Q_\perp^2$ is the square of 
the momentum transfer transverse to the to the magnetic field in the 
reactions~(\ref{1}) and~(\ref{3})
[with the corresponding substitutions $p \to - p$ and
$k \to - k$ in the crossing reactions~(\ref{2}) and~(\ref{4})], 
$\tilde\varphi_{\a\be} = \tilde F_{\a\be} / B$ is the
dimensionless dual magnetic-field tensor, $\sigma=\pm 1$ is the projection
of the proton spin on the direction of the magnetic field, $n$ is the
summation index over the proton Landay levels, 
${\cal T} V = {\cal T} L_x L_y L_z$ is the normalization four-volume,
$g_a$ is the axial constant of the nucleonic current, $G_F$ is the Fermi
constant, and $\theta_c$ is the Cabibbo angle. We note that in the limit
of a strong magnetic field, when the protons occupy only the ground
Landay level, expressions~(\ref{s2}) and~(\ref{m2}) agree with the
result obtained previosly in Ref.~\cite{Leinson}. 

Analysis shows that URCA processes are the fastest reactions in medium
considered, and they transfer the medium into a state of $\beta$ equilibrium 
in a time of order $10^{-2}$~s. Therefore we employed the condition of $\beta$
equilibrium and singled out in the expression for the energy-momentum
transfer to the shell (\ref{dpa}) the separate contributions from
processes involving neutrinos (\ref{1}), (\ref{3}) and antineutrinos
(\ref{2}), (\ref{4}):
\b
\frac{dP_\a^{\n}}{dt} = \int \frac{d^3k}{(2\pi)^3}\;\; k_\a \;
\left [ 1 +
\exp \left( \frac{- \omega + \mu_{\n}}{T} \right ) 
\right ] \; {\cal K}^{\n}  \, \delta f^{\n} \; .      \l{dpa1}
\e
Here, $\delta f^{\n}$ is the deviation of the distribution function from
the equilibrium  function, $ {\cal K}^{\n} $ is the (anti)neutrino
absorption coefficient, defined as
\b
{\cal K}^{\n} = \int \prod\limits_i dn_i f_i \prod\limits_f dn_f
(1 -f_f) \frac{|S_{if}|^2}{{\cal T}} \; , \l{k_n}
\e
where the integration extends over all states except the neutrino states
in the reaction (\ref{3}) and antineutrino states in the reaction
(\ref{4}), respectively. As follows from Eq.~(\ref{dpa1}), actually, the
momentum transferred to the medium is different from zero only if the
neutrino distribution function deviates from the equilibrium
distribution.

To calculate the absorption coefficient $ {\cal K}^{\n} $ we assumed that
the ultrarelativistic \ep plasma occupies only the ground Landau level,
while the protons occupy quite many levels (the dimensionless parameter
$ \delta = eB / m_p T \ll 1$). We also used
the fact that at the densities under consideration the nucleonic gas is
Boltzmannian and nonrelativistic. Then, dropping terms $\sim \delta$,
we can write expression (\ref{k_n}) in the form
\b
{\cal K}^{\n} & = & \frac{G_F^2 \cos^2 \theta_c \; eB \; N_{(n,p)}}{2\pi}  
\Bigg [ (1 + 3 g_a^2) - (g_a^2 - 1) k_3 / \omega  \Bigg ] 
\nonumber \\
& \times & \left [ 1 + \exp 
\left( \frac{ \pm (\mu_e - \triangle) - \omega}{T} \right ) 
\right ]^{-1}
\; ,
\e
where $N_n$, $N_p$, and $m_n$, $m_p$ are the number densities and masses 
of the neutrons and protons, respectively, $\triangle = m_n - m_p $, and
$ \omega $ and $k_3$ are the neutrino energy and the neutrino momentum 
in the direction of the magnetic field, respectively.

For further calculations we employed the neutrino distribution function
in the model of a spherically symmetric collapse of a supernova in the
absence of a magnetic field \cite{Janka}. This is a quite good
approximation when the region occupied by the strong magnetic field is
smaller than or of the order of the neutrino mean-free path. By the strong 
field we mean the field in which \ep plasma occupies only the ground Landau
level: $eB  \gtrsim \mu_e^2$. In the model of Ref.~\cite{Bisnovat} the region
occupied by such a field is no greater than several kilometers in size,
and we estimate the neutrino mean-free path in this region as
\b
l_\nu \simeq 4 \, {\rm km} \, 
\left ( \frac{4.4 \times 10^{16} \, {\rm G}}{B} \right ) \, 
\left ( \frac{5 \times 10^{11} {\rm g/cm}^3}{\rho} \right ) \; .
\l{l_n}
\e
Therefore the magnetic field cannot strongly alter the neutrino
distribution function, and our approximation is quite correct.

As calculations of the components of the energy-momentum (\ref{dpa1})
transferred to the medium during neutrino reemission showed, the radial
force arising is much weaker than the gravitational force and cannot
greatly influence the mantle dynamics. However, the force acting in the
direction of the magnetic field can change quite rapidly the distribution
of the angular velocities in the region occupied by the strong magnetic
field. The density of this force can be represented as
\b
\Im_{\|}^{(tot)} = \Im_{\|}^{(\nu)} + \Im_{\|}^{(\tilde\nu)} =
{\cal N} \Bigg[ \Bigg( 3 \left< \mu^2 \right>_{\nu}-1 \Bigg) I(a) e^{-a}  +
\Bigg( 3 \left< \mu^2 \right>_{\tilde\nu}-1 \Bigg) I(-a) \Bigg]  ,
\l{F}
\e
\b
\left< \mu^2 \right> = \left( \int \mu^2 \;\omega \;f \; d^3k \right)
\cdot \left( \int \omega \;f \; d^3k \right)^{-1}, \nonumber
\e
where $ \mu $ is the cosine of the angle between the neutrino momentum
and the radial direction, $a = \mu_e / T$, and 
\b
I (a) = \int\limits_0^\infty \frac{y^3 \, dy}{{\rm e}^{y-a} + 1} \; . 
\nonumber
\e
In deriving Eq. (\ref{F}) we use the one-dimensional factorized neutrino
distribution function $f^{\n} = \phi^{\n}(\omega / T_\nu) \,
\Phi^{\n}(r,\mu)$~\cite{R}, where $T_\nu$ is the neutrino spectral 
temperature and $r$ is the distance from the core center. 
To estimate the force in the diffusion region we assumed that
$ T_\nu \simeq T $ and chose $ \phi^{\n}(\omega / T_\nu) =
\exp{ (- \omega / T_\nu) } $. We determined the dimensional parameter
$ {\cal N} $ in expression (\ref{F}) as
\b
{\cal N} & = & \frac{G_F^2 \cos^2 \theta_c}{ (2\pi)^3 } \, 
\frac{g_a^2 - 1}{3} \, eB \, T^4 \, N_N
\l{N} \\[2mm]
& \simeq & 4.5 \times 10^{20} \, \frac{{\rm dynes}}{{\rm cm}^3} \, 
\left( \frac{T}{5~{\rm MeV}} \right)^4 
\left( \frac{B}{4.4 \times 10^{16}~{\rm G}} \right) 
\left( \frac{\rho}{5 \times 10^{11}~{\rm g/cm}^3} \right) \; ,
\nonumber
\e
where $N_N = N_n + N_p$ is the total nucleon number density.

The force~(\ref{F}) was estimated numerically in the diffusion region of
the supernova atmosphere for typical (excluding the field) values of the
macroscopic parameters for this region: $T = 5$~MeV, 
$B = 4.4 \times 10^{16}$~G, $\rho = 5 \times 10^{11}$~g/cm$^3$. 
For these values $a \simeq 3$, $\left < \mu^2 \right >_\nu \simeq
\left < \mu^2 \right >_{\tilde \nu} \simeq 0.4$ (Ref.~\cite{Janka}), 
and the force density in the direction of the field can be estimated from
Eq. (\ref{F}) as
\b
\Im^{tot}_\| \simeq \Im^{\nu}_\| \simeq {\cal N} \; . \l{N1}
\e
We note that the angular acceleration produced by the torque exerted by
such a force is large enough to spin up the region of the mantle
containing a strong magnetic field to typical angular velocities of a fast 
pulsar (with the rotational period $P_0 \sim 10^{-2}$~s) in a characteristic 
time of order of a second. In our opinion, this result is of interest in 
itself and can serve as a basic for a number of applications. However, we 
shall give a qualitative discussion of only one possible manifestation of 
this result -- the effect of such a fast spin-up of the mantle on the further 
generation of the toroidal magnetic field. Indeed, if the modification of 
the gradient of
the angular velocities of the mantle is large, the toroidal magnetic
field in the mantle at subsequent times will vary according to a law that
is different from the linear law \cite{Bisnovat}. Analysis of the
equation governing the generation of a toroidal field with allowance for
the force (\ref{N1}), which is linear in this field, leads to the
conclusion that its growth in time is much faster (exponential) in the
quite small region in which the force acts ($eB \gtrsim \mu_e^2$).
However, the main source of the magnetic field energy, just as in the
case when there is no force, is the kinetic energy of the
rigid-body-rotating core. Thus the force (\ref{N1}) can lead to a
peculiar rearrangement of the region occupied by the strong field. 
Specifically, with virtually no change in energy, the magnetic field can
become concentrated in a narrower spatial region and can therefore have
in this region higher intensities, on average, than in the absence of the
spin-up force. We note that the effect under discussion can strongly
influence the mantle-shedding process and also the mechanism leading to
the formation of an anisotropic $ \gamma $-ray burst in the explosion of
a supernova with a rapidly rotating core \cite{W}. However, in order to
perform detailed calculations of the generation of a toroidal field, the
enhancement of the field due to the "neutrino spin-up", and the
effect of this enhancement on the indicated processes, it is necessary to
analyze the complete system of MHD equations. That analysis lies far outside 
the scope of the present work and is a subject of a separate investigation.
Qualitative estimates show that the magnetic field with the strength 
$B \sim 10^{17}$~G can be generated by the above-described mechanism 
in a small neighborhood (of order a kilometer) of a rapidly rotating core
with the period $P_0 \sim 5 \times 10^{-3}$~s, and this region decreases 
with increasing the period.

In summary, we have shown that URCA neutrino reemission processes can
produce, in the region of the mantle that is filled with a strong
toroidal magnetic field, angular accelerations which are sufficiently
large as to greatly influence the mechanism of further generation of the
field. Specifically, such rapid redistribution of the angular velocities
can enhance the field in a small neighborhood of the rigid-body-rotating
core of a remnant.

We are grateful to S. I. Blinnikov for fruitful discussions and for
assistance in formulating the problem and to N. V. Mikheev and
M. V. Chistyakov for helpful discussions. This work was partially
supported by INTAS (Grant No. 96-0659) and the Russian Foundation 
for Basic Research (Grant No. 98-02-16694).

\end{document}